\documentstyle[amssymb,prl,aps]{revtex}

\renewcommand{\theequation}{\thesection.\arabic{equation}}
\let\ssection=\section
\renewcommand{\section}{\setcounter{equation}{0}\ssection}
\newcommand{\contr}{\hskip.7mm 
                    \rule[.1mm]{2.3mm}{.2mm}\rule[.1mm]{.2mm}{2.mm}\hskip.7mm}
\newcommand{\hodge}{{}^\star}
\newcommand{\D}{{\rm d}}
\newcommand{\DD}{{\rm D}}
\begin{document}

\title{Maxwell's theory on a post-Riemannian spacetime\\
       and the equivalence principle}

\author{Roland A. Puntigam, Claus L\"ammerzahl\thanks{Permanent address:
        Fakult\"at f\"ur Physik, Universit\"at Konstanz,
        D-78434 Konstanz, Germany} and Friedrich W. Hehl}
\address{Institute for Theoretical Physics, University of Cologne,
         D-50923 K{\"o}ln, Germany}
\maketitle
\bigskip\bigskip
\begin{abstract}
  The form of Maxwell's theory is well known in the framework of
  general relativity, a fact that is related to the applicability of
  the principle of equivalence to electromagnetic phenomena. We pose
  the question whether this form changes if torsion and/or
  nonmetricity fields are allowed for in spacetime. Starting from the
  conservation laws of electric charge and magnetic flux, we recognize
  that the Maxwell equations themselves remain the same, but the
  constitutive law must depend on the metric and, additionally, may
  depend on quantities related to torsion and/or nonmetricity. We
  illustrate our results by putting an electric charge on top of a
  spherically symmetric exact solution of the metric-affine gauge theory of 
  gravity (comprising torsion and nonmetricity). All this is compared to the
  recent results of Vandyck.
\noindent\pacs{PACS no.:}
\end{abstract}
\noindent{\it file vandyck8.tex, 1997-02-11, Classical 
  and Quantum Gravity, to be published (1997)}\bigskip

\section{Introduction}

It was Minkowski, in 1908, who formulated Maxwell's theory in a
four-dimensional flat pseudo-Euclidean spacetime, Minkowski's
special-relativistic `world'. The next step, generalizing Maxwell's
theory if {\it gravity} can no longer be neglected, was performed by
Einstein and Grossmann in 1913. They `lifted' Maxwell's theory to a
four-dimensional pseudo-Riemannian spacetime. This amounted to a
successful application of the equivalence principle to Maxwell's
theory. Not too much later, after the creation of general relativity
theory, Einstein \cite{Einstein16} reformulated Maxwell's theory such
that it became apparent that the basic structure of Maxwell's theory,
namely the field equations, is left intact even when a metric is not
used.  Later Kottler \cite{Kottler}, E.~Cartan \cite{Cartan}, van
Dantzig \cite{Van}, and others put forward the so-called metric-free
formulation of electrodynamics, see Post \cite{Post62,Post80} and
Schouten \cite{Schouten89}. It was recognized in this general
framework that Maxwell's equations can be understood as arising from
the conservation laws of {\it electric charge} and {\it magnetic
  flux}, see Truesdell and Toupin \cite{Truesdell60}.

These conservation laws can be reduced to {\it counting statements},
since electric charge comes in quantized portions of elementary
charges (or rather as one thirds of them) and magnetic flux can also
exist, in superconducting media, in a quantized form, the flux quantum
or fluxoid $h/(2e)$, as was predicted (up to a factor 2) already by
F.~London \cite{London50}. Thus it is obvious that the formulation of
these laws only requires a four-dimensional {\it differentiable
  manifold} and the possibility of a foliation of it into
three-dimensional hypersurfaces.  The {\it constitutive laws} do need
a metric, in contrast to the Maxwellian field equations themselves, a
point of view which has been repeatedly stressed by Post
\cite{Post80,Post95}, see also Bamberg \& Sternberg \cite{Bamberg}.
Any {\it post}-Riemannian geometry, that is, any spacetime geometry
which has more geometrical field variables (`gravitational fields')
than the metric, is irrelevant to the Maxwell equations. In
particular, neither torsion nor nonmetricity couple to it, a point
which has already been made by Benn, Dereli, and Tucker \cite{Benn}. 
Only  the constitutive law may depend in a very restricted
way on torsion and nonmetricity structures.

In the Einstein-Cartan theory of gravity, spacetime carries an additional 
{\it torsion}, and, still more general, in the framework of metric-affine 
gravity \cite{Hehl95}, a {\it nonmetricity} enters the geometrical arena 
of spacetime. What can we predict about Maxwell's theory under those more 
exotic circumstances? Can we again apply the equivalence principle? Should 
we write down Maxwell's equations in the Minkowski world in Cartesian 
coordinates and replace the partial by covariant derivatives, or should 
we do that in the Lagrangian? What type of coupling to gravity should we 
assume? 

Especially the application of the `comma goes to semicolon rule' (see
Misner, Thorne, and Wheeler \cite{MTW}) creates difficulties for
charge conservation if applied to the inhomogeneous Maxwell equation
in post-Riemannian spacetimes. Some test theory for the coupling of
the Maxwell equations to non-metric structures of spacetime has been
investigated by Coley \cite{Coley}.

Vandyck has addressed these questions in a recent article
\cite{Vandyck96}. We find his answers not totally convincing.  Therefore
we will try to argue that the axiomatic formulation of Maxwell's
theory, alluded to above, is sufficient for formulating Maxwell's
theory in such post-Riemannian spacetimes, including possibly torsion
and nonmetricity.

As formalism we use exterior calculus, for our conventions see \cite{Hehl95}.

\section{Electric charge conservation}\label{sec:charge}

Let be given the odd (or twisted, see \cite{Burke85}) electric current
three-form $J$. We assume that a $(1+3)$-foliation of spacetime holds 
locally. The different three-dimensional hypersurfaces are labeled
by a parameter $\tau$. We introduce a normal vector $n$ such that
$n\contr\D\tau = 1$. Then we can decompose the current three-form 
according to 
\begin{equation}
J=\rho-j\wedge\D\tau\,,\label{current}
\end{equation}
where $\rho$ is the charge density three-form and $j$ the electric
current two-form. As axiom 1 we assume electric charge conservation 
($\D$ = four-dimensional exterior derivative):
\begin{equation}\label{chargecon}
   \oint_{\partial V_4}J = \int_{V_4}\D J = 0 \qquad\hbox{(Axiom 1)}\,. 
\end{equation}
Here $V_4$ is an arbitrary four-dimensional volume and $\partial V_4$ its 
three-dimensional boundary. If this is assumed to be valid for all 
three-cycles $c_3=\partial V_4$, then $J$ is exact \cite{Post95,Post79,Post82}:
\begin{equation}\label{Maxwellinhom}
   \fbox{\D G = J\,.}
\end{equation}
The electromagnetic {\it excitation} $G$ is an odd two-form which
decomposes as 
\begin{equation}
   G = D-H\wedge\D\tau\,.\label{excitation}
\end{equation}
Therefore (\ref{Maxwellinhom}) is equivalent to the inhomogeneous
Maxwell equations
\begin{eqnarray}
  \underbar{d}D&=&\rho\qquad\hbox{(Gauss law)}\,,\label{Gauss}\\ 
  \underbar{d}H-\dot{D} &=& j\qquad\hbox{(Oersted-Amp\`ere law)}
  \label{Oersted}\,;
\end{eqnarray}
here $\underbar{d}:=\D-\D\tau\,( n\contr \D)\wedge$ is the
three-dimensional exterior derivative, and $\dot D := \pounds_n D$ is
defined via the Lie derivative, for details compare \cite{Hehl91b} and
the literature given. Note that, up to now, only the differential
structure of the spacetime was needed.  The electric excitation $D$
can be measured by means of Maxwellian double plates (see Pohl
\cite{Pohl}) as charge per unit area, the magnetic excitation by means
of a small test coil, which compensates the $H$-field to be measured,
as current per unit length. In other words, the extensive quantities
$D$ and $H$ have an operational significance provided we know the
characterizing properties of an ideal conductor.

\section{Lorentz force}

{}From mechanics we take the notion of an even covector-valued force
density four-form $f_\alpha$. In the conventional manner, we {\it define} 
the electromagnetic {\em field strength} $F$ via axiom 2:
\begin{equation}\label{Lorentz}
   f_\alpha = \left(e_\alpha\contr F\right)\wedge J\qquad\hbox{(Axiom 2)}\,.
\end{equation}
{}From mechanics originates the notion of $f_\alpha$, from axiom 1 the
current $J$, the $e_\alpha$'s denote the frame. The even
two-form $F$ can be decomposed as
\begin{equation}\label{Fdecomp}
   F = B+E\wedge\D\tau\,,
\end{equation}
that is, for $a,b=1,2,3$,
\begin{equation}\label{3force}
   {\frak f}_a = -\rho\left(e_a\contr E\right)-j\wedge\left(e_a\contr B\right)
   \qquad\mbox{with}\qquad 
   f_\alpha = {\frak f}_\alpha \wedge\D \tau\,.
\end{equation}
Therefore the Lorentz force (\ref{3force}), via (\ref{Fdecomp}),
yields an operational definition of the electromagnetic field strength
$F$ as a force field -- and hence as an intensive quantity. Again no
metric nor a connection is necessary for formulating axiom 2. 

An alternative way of introducing $F$ -- again metric etc.\ 
independent -- is provided by quantum interference measurements of the
Aharonov-Bohm type yielding an observable phase shift
$\delta\varphi=\frac{e}{\hbar}\int_{V_2}F$.

Now we have to impose some conditions on the newly defined field
strength $F$.

\section{Magnetic flux conservation}

The field strength $F$ is a two-form. Thus we can postulate the
conservation of magnetic flux as axiom 3:
\begin{equation}\label{fluxcon}
   \oint_{\partial V_3}F = \int_{V_3}\D F = 0\qquad\hbox{(Axiom 3)}\,.
\end{equation}
By Stokes' theorem and the arbitrariness of the two-cycles $c_2=\partial
V_3$, we have
\begin{equation}\label{Maxwellhom}
   \fbox{\D F = 0\,.}
\end{equation}
In $(1+3)$-decomposition this reads 
\begin{eqnarray}
   \underbar{d}B&=&0\qquad\hbox{(magnetic field closed)}\,,\label{dB=0}\\
   \underbar{d}E+\dot{B}&=&0\qquad\hbox{(Faraday law)}\,.\label{Faraday}
\end{eqnarray}

Maxwell's equations are represented by (\ref{Maxwellinhom},
\ref{Maxwellhom}) or, equivalently, by (\ref{Gauss}, \ref{Oersted},
\ref{dB=0}, \ref{Faraday}). In this form, they are generally
covariant, i.e.\ valid in arbitrary frames and arbitrary coordinates.
Moreover, neither metric nor torsion or nonmetricity take part in this
set-up. Therefore, if one starts from a four-dimensional differential
manifold, which admits a $(1+3)$-foliation, and introduces a metric
and a connection, then the structure of the Maxwell equations
(\ref{Maxwellinhom}) and (\ref{Maxwellhom}) is insensitive to it and
does {\it not} change.

We can argue similarly in the complementary situation: In a Minkowski
space, we formulate the Maxwell equations in the way we did. Then they
keep their form with respect to an {\it accelerated} frame.
Consequently, switching on {\it gravity} and requiring the equivalence
principle to be valid, the Maxwell equations must not change
either. In spite of the `deformation' of spacetime by means of
gravity, the Maxwell equations remain `stable'. Therefore, in this
framework of deriving Maxwell's theory from electric charge and
magnetic flux conservation, the Maxwell equations stay the same ones
in a Minkowskian, a Riemannian, or a post-Riemannian spacetime. No
additional effort is needed in order to adapt the Maxwell equations if
a spacetime is considered with additional geometrical attributes.
This is the most straightforward application of the equivalence
principle one can think of.

\section{Constitutive law}\label{sec:const}

So far, the Maxwell equations (\ref{Maxwellinhom}) and
(\ref{Maxwellhom}) represent an underdetermined system of evolution
equations for $G$ and $F$.  In order to reduce the number of
independent variables, we have to set up a relation between $G$ and
$F$:
\begin{equation}
G = G(F) \,. \label{clgen}
\end{equation}
Special cases of this constitutive law are:

\begin{description}
\item[(i)] {\it Vacuum:} The standard constitutive law for vacuum is
\begin{equation}
  G = \hodge F \,.\label{cl1}
\end{equation}
On the right-hand side of (\ref{cl1}), the factor
$(\varepsilon_0/\mu_0)^{1/2}$ has been absorbed for simplicity.  Here, by
means of the Hodge star, the metric enters the Maxwell theory for the
first time. The appearance of the metric is necessary from a physical
point of view in order to get the {\it light cone} as characteristic
surface of the evolution equations for the Maxwellian field strength
$F$. The law (\ref{cl1}) is valid in Minkowksi, Riemannian, and
post-Riemannian spacetimes likewise.

\item[(ii)] {\it Axion:} The constitutive law of the vacuum
  (\ref{cl1}) relates the ordinary two-form $F$, via the Hodge star (which is 
  twisted), to the twisted two-form $G$. If we had a twisted zero-form 
  $\theta$ (`pseudo-scalar') at our disposal, then we could supplement
  the right hand side of the vacuum law by the twisted term $\theta\,F$:
  \begin{equation}\label{axion}
     G= \hodge F+\theta\,F= (\,\hodge+\theta)\,F\,.
  \end{equation} 
  The exterior derivative of this equation, because of $\D F=0$, 
  turns out to be
  \begin{equation}\label{dGaxion}
    \D G= \D\,\hodge F+\D\theta\wedge F\,.
  \end{equation}
Thus the inhomogeneous Maxwell equation, in terms of $F$, reads
\begin{equation}\label{daxion}
  \Bigl(\D\,\hodge +(\D\theta)\wedge\Bigr)\, F=J\,,\quad\text{with}\quad
  \D J=0\,.
\end{equation}
The `pseudo-scalar' field $\theta$ is known in the literature as
hypothetical axion field, see \cite{Weinberg} and \cite{Wilczek}. Its
possible implications for cosmology are discussed in \cite{Kolb}. The
axion-Maxwell interaction Lagrangian turns out to be
$\sim\theta\,F\wedge F=\theta\,\D(F\wedge A)$.

We can relate the axion field to the torsion of spacetime. The torsion
$T^\alpha$ is an ordinary two-form. Its axial piece is proportional to
the ordinary three-form $T^\alpha\wedge\vartheta_\alpha$. The dual of it
is a twisted one-form $\hodge(T^\alpha\wedge\vartheta_\alpha)$.
Therefore, with some constant $c$, we can make the identification
\begin{equation}\label{axionansatz}
  \D\theta=c\,\hodge(T^\alpha\wedge\vartheta_\alpha)\,,\end{equation}
which yields (Gasperini and de Sabbata \cite{Gasperini}, see also
\cite{Venzo})
\begin{equation}\label{axiontorsion}
  \Bigl(\D\,\hodge\,+c\,\hodge (T^\alpha\wedge\vartheta_\alpha)\wedge
  \Bigr)\,F=J\,.
\end{equation} 
This equation describes the coupling of the inhomogeneous Maxwell equation 
to an axial piece of the torsion.  However, this interpretation is not 
compulsory. Incidentally, Eq.(\ref{axiontorsion}) seems to represent the most 
general post-Riemannian coupling linear in $F$ which is compatible with charge
conservation \cite{Lammerzahl97}; a piece with the Weyl covector, e.g., is
excluded since it is an ordinary, but not a twisted form. 

We recognize also in this example that there doesn't seem to exist
a chance to introduce other post-Riemannian structures in the
axion-Maxwell equation (\ref{axiontorsion}) in an ad hoc way.
We would like to stress that (\ref{axiontorsion}) is valid in a 
spacetime with arbitrary metric and connection.

\item[(iii)] {\it Born-Infeld:} The non-linear Born-Infeld theory
  \cite{BI} represents a classical generalization of Maxwell's theory
  for accommodating stable solutions for the description of
  `electrons'.  Its constitutive law reads (with a dimensionful
  parameter $f$, the so-called maximal field strength, see also
  \cite{GiRa}):
\begin{equation}
  G = \frac{\hodge F - \frac{1}{2f^2}\, \hodge(F\wedge F)\, F
    }{\sqrt{1 + \frac{1}{f^2}\,\hodge(F\wedge\hodge F) -
      \frac{1}{4f^4}\,[\hodge(F\wedge F)]^2}}\,.
\end{equation}
It leads to a non-linear equation for the dynamical evolution of the
field strength $F$. As a consequence, the characteristic surface, the
light cone, depends on the field strength, and the superposition
principle for the electromagnetic field doesn't hold any longer.

\item[(iv)] {\it Heisenberg-Euler:\/} Quantum electrodynamical vacuum
  corrections to Maxwell's theory can be accounted for by an effective
  constitutive law constructed by Heisenberg and Euler \cite{HE}. To second 
  order in the fine structure constant $\alpha$, it is given by 
  (see also \cite{Itzykson})
\begin{equation}
  G = \Bigr[1 + \frac{4\,\alpha^2}{45\, m^4}\, \hodge\bigl(F\wedge\hodge F
      \bigr)\Bigr]\, \hodge F 
    + \frac{7\,\alpha^2}{45\,m^4} \,\hodge\bigl(F\wedge F\bigr)\,F\,,
\end{equation}
where $m$ is the mass of the electron. Again, post-Riemannian
structures don't interfere here.
\end{description}

\section{Energy-momentum current of the electromagnetic field}

For quantifying the gravitational effect of the electromagnetic field, we
need its energy-momentum current. The Lagrangian four-form of Maxwell's 
field reads
\begin{equation}\label{maxenergy}
   L_{\rm Max} = - \frac{1}{2} F \wedge G \,. 
\end{equation}
The canonical energy-momentum current is computed from the Lagrangian 
four-form (\ref{maxenergy}) and can be represented by the odd covector-valued
three-form
\begin{equation}\label{momentum}
   \Sigma^{\rm Max}_\alpha = e_\alpha \contr L_{\rm Max} + (e_\alpha \contr F)
   \wedge G = \frac{1}{2}\left[(e_\alpha\contr F)\wedge G 
   - (e_\alpha\contr G)\wedge F \right]\,.
\end{equation}
This energy-momentum current will enter the right hand side of the first
field equation, as we will see below.

\section{An Electric Charge in Einstein-Dilation-Shear Gravity}
\label{sec:mag}

As a nontrivial example, let us consider the electromagnetic field in
the framework of the metric-affine gauge theory (MAG) of gravity
\cite{Hehl95}, in particular its effect on an exact solution of this
theory \cite{Obukhov96b}, see also \cite{Vlachynsky96}. Similar 
solutions have been found by Tucker and Wang, see \cite{Tucker1} 
and \cite{Tucker2}.  The geometrical
ingredients of MAG are the curvature two-form $R_\alpha{}^\beta =
\frac{1}{2}\,R_{ij\alpha}{}^\beta \D x^i\wedge \D x^j$, and, as
post-Riemannian structures, the nonmetricity one-form
$Q_{\alpha\beta}=Q_{i\alpha\beta}\,\D x^i$ and the torsion two-form
$T^\alpha=\frac{1}{2}\,T_{ij}{}^\alpha \D x^i\wedge \D x^j$. The
simple toy model that we want to consider is specified by a
gravitational gauge Lagrangian, quadratic in curvature, torsion, and
nonmetricity,  see \cite{Obukhov96b},
\begin{equation}\label{Vdil-sh}
  V_{\text{dil-sh}}=-{1\over2\kappa}\,\left(R^{\alpha\beta}\wedge
    \eta_{\alpha\beta} - 2 \lambda \eta +
    \beta\,Q\wedge\hodge Q+\gamma\,T\wedge\hodge T\right)
  -{\alpha\over 8}\,R_\alpha{}^\alpha\wedge\hodge R_\beta{}^\beta\,,
\end{equation}
coupled to the Maxwell Lagrangian (\ref{maxenergy}) according
to $L_{\rm tot}=V_{\text{dil-sh}}+L_{\rm Max}$. In (\ref{Vdil-sh}) 
we have introduced the Weyl covector $Q := Q_\gamma{}^\gamma/4$ and
the covector piece of the torsion $T := e_{\alpha}\contr T^{\alpha}$.
Einstein's gravitational constant is denoted by $\kappa=\ell^2/(\hbar c)$ 
(with the Planck length $\ell$), and $\lambda$ is the cosmological constant.
The coupling constants $\alpha, \beta$, and $\gamma$ are dimensionless.

Varying the coframe and the connection, we find the two relevant field 
equations of MAG \cite{Hehl95}, 
\begin{eqnarray} 
   \DD H_{\alpha} - E_{\alpha} &=& \Sigma_{\alpha}\,,\label{first}\\ 
   \DD H^{\alpha}{}_{\beta}-E^{\alpha}{}_{\beta}&=&\Delta^{\alpha}{}_{\beta}\,,
   \label{second}
\end{eqnarray}
referred to as the {\em first} and the {\em second} field equation, 
respectively, with $\DD$ as the covariant exterior derivative. 
In (\ref{first}) and (\ref{second}) there enter the canonical 
energy-momentum and hypermomentum currents of matter $\Sigma_{\alpha}$ and 
$\Delta^{\alpha}{}_{\beta}$, the gravitational gauge field momenta 
\begin{equation}
   H_{\alpha} := -{\partial V_{\text{dil-sh}}\over \partial T^{\alpha}} 
   \quad\mbox{and}\quad 
   H^{\alpha}{}_{\beta} := - {\partial V_{\text{dil-sh}}\over \partial
   R_{\alpha}{}^{\beta}} \,,
\end{equation}
and the canonical energy-mo\-men\-tum and hypermomentum currents of the 
gauge fields
\begin{eqnarray} 
   E_{\alpha} & =&  e_{\alpha}\contr V_{\text{dil-sh}} 
   + (e_{\alpha}\contr T^{\beta}) \wedge H_{\beta} 
   + (e_{\alpha}\contr R_{\beta}{}^{\gamma})\wedge H^{\beta}{}_{\gamma} 
   + {1\over 2}(e_{\alpha}\contr Q_{\beta\gamma}) M^{\beta\gamma}\,,\\
   E^{\alpha}{}_{\beta} & =&  
   - \vartheta^{\alpha}\wedge H_{\beta} - M^{\alpha}{}_{\beta} \,.
\end{eqnarray} 
The gravitational gauge field momentum $M^{\alpha\beta}$ is coupled to 
the nonmetricity:
\begin{equation}
   M^{\alpha\beta} := 
   -2\,{\partial V_{\text{dil-sh}}\over \partial Q_{\alpha\beta}}\,.
\end{equation}
We study only the behavior of the electromagnetic field in the
metric-affine framework. Thus, for the matter currents in (\ref{first}) and 
(\ref{second}) we have $\Sigma_\alpha = \Sigma^{\rm Max}_\alpha$, 
cf.~(\ref{momentum}), and $\Delta^{\alpha}{}_{\beta} = 0$.  

The formalism of MAG, as outlined in the present section, is not limited to 
the simple and very restricted Lagrangian (\ref{Vdil-sh}); more general 
choices for the Lagrangian are possible. It is our intention, however, not to 
look at MAG for its own interest, but to investigate the behavior of Maxwell's 
theory within a non-Riemannian spacetime. This was our motivation for making
the simplest possible choice of the metric-affine part of the Lagrangian that 
still allows for propagating torsion and nonmetricity, see 
Obukhov et al.~\cite{Obukhov96b}.

\section{Exact Solution with Spherical Symmetry}\label{sec:sol}
  
The field equations (\ref{first}) and (\ref{second}), together with
Maxwell's equations (\ref{Maxwellinhom}) and (\ref{Maxwellhom}) -- assuming
the constitutive law (\ref{cl1}) -- are approached as follows. The 
spherically symmetric coframe
\begin{equation}\label{frame2}
  \vartheta ^{\hat{0}} =\, f\, \D t \,,\quad
  \vartheta ^{\hat{1}} =\,{1\over f}\, \D r\,, \quad
  \vartheta ^{\hat{2}} =\, r\,\D\theta\,,\quad
  \vartheta ^{\hat{3}} =\, r\, \sin\theta \, \D\phi\,,
\end{equation}
contains the zero-form $f = f(r)$ and is assumed to be orthonormal, i.e., 
the metric reads
\begin{equation}\label{schwarz}
   ds^2 = o_{\alpha\beta}\,\vartheta^\alpha\otimes\vartheta^\beta 
        = -f^2\,\D t^2 + \frac{1}{f^2}\D r^2
          +r^2\left(\D\theta^2+\sin^2\theta \,\D\phi^2\right)\,.
\end{equation}
The nonmetricity one-form is taken to contain only two irreducible pieces
(see \cite[Appendix B.1]{Hehl95}),
\begin{equation}\label{QQ}
  Q_{\alpha\beta}=\, ^{(3)}Q_{\alpha\beta} +\,
  ^{(4)}Q_{\alpha\beta}\,,
\end{equation}
namely the dilation (or Weyl) piece $^{(4)}Q_{\alpha\beta} 
= Q g_{\alpha\beta}$ and a proper shear piece
\begin{equation}\label{3q}
   ^{(3)}Q_{\alpha\beta} = {4\over9}\left(\vartheta_{(\alpha}e_{\beta)}
   \contr \Lambda 
   - {1\over4}g_{\alpha\beta}\Lambda\right)\,,\qquad \hbox{with}\qquad
   \Lambda:= \vartheta^{\alpha}e^{\beta}\contr\!
   {\nearrow\!\!\!\!\!\!\!Q}_{\alpha\beta} \,.
\end{equation}
Furthermore, we allow only the covector piece ${}^{(2)}T^{\alpha}$ in the 
torsion two-form:  
\begin{equation}
   T^{\alpha} = {}^{(2)}T^{\alpha}={1\over 3}\,\vartheta^{\alpha}\wedge T\,.
\end{equation}
Finally, we use a spherically symmetric electric (Coulomb) charge at the
origin of the spatial coordinates with the corresponding field strength
\begin{equation}
   F = \frac{q}{r^2}\, \vartheta^{\hat 1} \wedge \vartheta^{\hat 0} \,,
\end{equation}
and we impose the constitutive law (\ref{cl1}).

With these prescriptions and the ansatz
\begin{equation}\label{genEug}
   Q=u(r)\,\vartheta^{\hat{0}}\,,\quad\quad\Lambda=v(r)\,\vartheta^{\hat{0}}\,,
   \quad\quad T=\tau(r)\,\vartheta^{\hat{0}}
\end{equation}
for the {\em one-form triplet} $(Q, \Lambda, T)$, the solution is expressed by
\begin{equation}\label{f3}
   f = \sqrt{1 - \frac{2\kappa M}{r} + \frac{\lambda\,r^2}{3} 
     + \frac{\kappa\, q^2}{2 r^2} 
     + \alpha\frac{\kappa \widetilde{N}\rule{.pt}{12pt}^2}{2 r^2}}
\end{equation}
and
\begin{equation}\label{coul1}
   u = \frac{\widetilde{N}}{f r}\,,\qquad 
   v = \frac{3\beta}{2}\; \frac{\widetilde{N}}{f r}\,,\qquad 
  \tau = -\frac{\beta+6}{4}\,\frac{\widetilde{N}}{f r}\,,
\end{equation}
where $\widetilde{N}$ is an integration constant.
The dimensionless coupling constants are subject to the {\em constraint}
\begin{equation}\label{constraint}
   \gamma=-\frac{8}{3}\;\frac{\beta}{\beta+6}\,,
\end{equation}
i.e., only two of the post-Riemannian coupling constants 
$(\alpha,\beta,\gamma)$ in (\ref{Vdil-sh}) remain independent, 
while the third one, $\gamma$, is determined by (\ref{constraint}).
These results have been found with the help of the computer algebra system
REDUCE~\cite{Hearn93} making also use of its Excalc 
package~\cite{Schrufer94}, see \cite{Puntigam95sa}. 

Let us summarize the properties of the MAG-Maxwell solution that is
presented here. The zero-form $f$, that fixes the orthonormal coframe
(\ref{frame2}), has four contributions, see (\ref{f3}). The terms
containing the mass parameter $M$, the cosmological constant $\lambda$,
and the electric charge $q$ correspond exactly to the (general relativistic) 
Reissner-Nordstr\"om solution with cosmological constant. The additional 
term with the dilation charge $\widetilde{N}$ has a similar structure as
the previous term with the electric charge $q$. 
The nonmetricity has the explicit form 
\begin{equation}\label{nichtmetrizitaet}
   Q^{\alpha\beta} = \frac{\widetilde{N}}{f r}\,\left[o^{\alpha\beta}
   + \frac{2}{3}\,\beta\left(\vartheta^{(\alpha}e^{\beta )}\contr
   -\frac{1}{4}\,o^{\alpha\beta}\right)\right]\vartheta^{\hat{0}}\,
\end{equation}
carrying, besides the dilation piece, a shear part -- the second term in 
(\ref{nichtmetrizitaet}) with the factor $\beta$. The torsion two-form 
evaluates to 
\begin{equation}\label{tosion}
   T^\alpha = -\frac{\beta+6}{12}\,
   \frac{\widetilde{N}}{fr}\,\,\vartheta^\alpha\wedge\vartheta^{\hat{0}}\,,
\end{equation}
and the Faraday two-form 
\begin{equation}\label{FF}
   F = \frac{q}{r^2}\, \vartheta^{\hat 1} \wedge \vartheta^{\hat 0}
\end{equation}
has the same innocent appearance as that of a point charge in {\em
  flat} Minkowski space. It is clear, however, that all relevant
geometric objects, coframe, connection, torsion, curvature, etc.,
`feel' -- via the zero-form $f$ -- the presence of the electric
charge. However, as one can recognize from (\ref{nichtmetrizitaet},
\ref{tosion},\ref{FF}), the Maxwell field is disconnected from
nonmetricity and torsion otherwise. This exemplifies and is in full 
accordance with our general statement concerning the coupling of the 
Maxwell equations to post-Riemannian structures.

\section{Discussion}

There is so much experimental evidence in favor of the conservation
laws of electric charge and magnetic flux that one can hardly doubt
the correctness of axiom 1 and axiom 3 from a physical point of view.
If so, then the form of the Maxwell equations is fixed, and we have no
trouble in predicting how they change in spacetimes with Riemannian
and post-Riemannian geometrical structure: {\it They don't change at
  all}. They are stable against such `deformations'. Thereby the
equivalence principle turns out to be rather trivial in this context.
The only `freedom' one has is to modify the constitutive law. 
Incidentally, if the limits of classical physics are reached, then, 
on the level of quantum mechanics, a fresh look at the equivalence 
principle is needed, see \cite{Lammerzahl96}.

Coming back to the article of Vandyck \cite{Vandyck96}, we recognize
that the different options for generalizing the Maxwell equations are
artificial ones in the sense that they violate the well-established
axioms 1 and 3, namely the conservation of electric charge and
magnetic flux. These options can only emerge, if one forgets the
underlying physical structure of Maxwell's theory. Clearly, whether
one uses the calculus of tensor analysis (see the Appendix) or that of
exterior differential forms, doesn't make any difference, if one
starts off with our axioms.

In the framework of the Poincar\'e gauge theory of gravitation, the
spacetime of which carries, besides the metric, a propagating torsion,
we also found exact electrically charged solutions, see
\cite{Baekler88a}.  In this latter context, as well as in the case of
the new charged solution of MAG that was presented in
Sect.~\ref{sec:sol}, we used Maxwell's theory as described in
Sects.~\ref{sec:charge} to \ref{sec:const}.  And everything is
well-behaved and consistent with our analysis of how to couple the
Maxwell equations to post-Riemannian structures. There is almost no
freedom for an alternative coupling of Maxwell's equations to gravity
within Riemannian or post-Riemannian spacetimes. The equations
(\ref{Maxwellinhom}), (\ref{Maxwellhom}), and (\ref{cl1}) solve the
problem completely.

\section*{Appendix: The tensor analysis version of metric-free electrodynamics}
\renewcommand{\theequation}{A.\arabic{equation}}

We decompose excitation, field strength, and current into (holonomic)
coordinate components:
\begin{equation}
   G = \frac{1}{2!}\, G_{ij}\, \D x^i \wedge \D x^j\,,\quad
   F = \frac{1}{2!}\, F_{ij}\, \D x^i \wedge \D x^j\,,\quad
   J = \frac{1}{3!}\, J_{ijk}\, \D x^i \wedge \D x^j \wedge \D x^k\,.
\end{equation}
If we use the Levi-Civita antisymmetric unit tensor {\em density}
$\epsilon^{ijkl} = \pm 1,0$, which is metric-free, 
\begin{equation}
   {\cal G}^{ij} := \frac{1}{2!}\, \epsilon^{ijkl}\, G_{kl}\,, \quad
   {\cal J}^{i} := \frac{1}{3!}\, \epsilon^{ijkl}\, J_{jkl}\,,
\end{equation}
then Maxwell's equations read
\begin{equation}\label{maxcomp} 
   \partial_k {\cal G}^{ik} = {\cal J}^{i}\,, \qquad
   \partial_{[i} F_{jk]} = 0\,.
\end{equation}
The constitutive law for vacuum can be put in the linear form
\begin{equation}
  {\cal G}^{ij} = \frac{1}{2}\,\chi^{ijkl} F_{kl}\,, \qquad
  \chi^{(ij)kl} = \chi^{ij(kl)}=\chi^{[ijkl]} = 0\,,\quad
\chi^{ijkl}=\chi^{klij}\,,
\end{equation}
with the specific metric dependent ``modulus''
\begin{equation}\label{vaccomp}
   \chi^{ijkl} := 2 \sqrt{|\mbox{det}\,g_{mn}|}\; g^{k[i} g^{j]l} \,. 
\end{equation}
Eqs.~(\ref{maxcomp}) to (\ref{vaccomp}) are unchangedly valid in
post-Riemannian spacetimes. Note that the representation of the 
electromagnetic excitation ${\cal G}^{ij}$ as density, see 
Schr\"odinger \cite{Schrodinger54}, is vital for these considerations 
and distinguishes our approach from that of Vandyck.

\begin{acknowledgements}
  We are grateful to Werner Esser, Eckehard Mielke, Yuri Obukhov, and
  Norbert Straumann (Z\"urich) for helpful remarks. The first named
  author (RAP) is supported by the {\em Graduiertenkolleg Scientific
    Computing}, Cologne-St.Augustin, the second named author (CL)
  thanks the Deutsche Forschungsgemeinschaft, Bonn for financial
  support.
\end{acknowledgements}

\end{document}